\documentclass[%
aip,
amsmath,amssymb,
reprint,%
author-numerical,%
]{revtex4-2}

\usepackage{graphicx}
\usepackage{dcolumn}
\usepackage{bm}

\usepackage[utf8]{inputenc}
\usepackage[T1]{fontenc}
\usepackage{mathptmx}
\usepackage{etoolbox}

\usepackage{longtable}
\usepackage{rotating}

\usepackage{tabularx}
\usepackage{siunitx}
\usepackage{amssymb}
\usepackage{amsmath}
\usepackage{physics}
\usepackage{verbatim}
\usepackage{makecell}
\usepackage{bm}
\usepackage{float}
\usepackage{soul}
\usepackage{xcolor}
\usepackage{hyperref}
\usepackage{csquotes}
\sethlcolor{yellow} 
\makeatletter
\def\@email#1#2{%
 \endgroup
 \patchcmd{\titleblock@produce}
  {\frontmatter@RRAPformat}
  {\frontmatter@RRAPformat{\produce@RRAP{*#1\href{mailto:#2}{#2}}}\frontmatter@RRAPformat}
  {}{}
}%
\makeatother

\begin{document}

\preprint{AIP/123-QED}

\title[Continuing progress toward fusion energy breakeven and gain as measured against the Lawson criteria]%
{Continuing progress toward fusion energy breakeven and gain as measured against the Lawson criteria}

\author{Samuel E. Wurzel}
\affiliation{Fusion Energy Base, New York, NY 10003, USA}
\email{sam@fusionenergybase.com}

\author{Scott C. Hsu}
\affiliation{Lowercarbon Capital, Los Angeles, CA 90034, USA}

\date{\today}

\begin{abstract}
This paper is an update to our earlier paper ``Progress toward fusion energy breakeven and gain as measured against the Lawson criterion'' [Phys.\ Plasmas~{\bf 29}, 062103 (2022)]. Plots of Lawson parameter and triple product vs.\ ion temperature and triple product vs.\ date achieved are updated with recently published experimental results. A new plot of scientific energy gain vs.\ date achieved is included. Additionally, notes on new experimental results, clarifications, and a correction are included. 
\end{abstract}

\maketitle

\section{\label{sec:level1}Introduction}
This paper is an update to our earlier paper ``Progress toward fusion energy breakeven and gain as measured against the Lawson criterion,''\cite{2022_Wurzel_Hsu} which we refer to hereafter as the ``original paper.'' Since its publication in 2022, both fusion researchers and investors have encouraged us to provide updated plots and data tables highlighting the continued physics progress of fusion experiments. 

The purpose of this paper is therefore to update the set of experimentally achieved physics results with data published in peer‐reviewed publications since the beginning of 2022. These include results from the C-2W experiment (TAE Technologies), the Fusion Z-pinch Experiment (FuZE) (Zap Energy), the Joint European Torus (JET) (UK Atomic Energy Authority),  the Magnetized Liner Inertial Fusion (MagLIF) platform at the Z Facility (Sandia National Laboratories), the National Ignition Facility (NIF) (Lawrence Livermore National Laboratory), the OMEGA Laser Facility (University of Rochester’s Laboratory for Laser Energetics), the Plasma Compression System (PCS) and Plasma Injector 3 (PI3) experiments (General Fusion), and the ST40 spherical tokamak (Tokamak Energy).
Also included is a clarification of nomenclature, a slight reframing of the ICF (inertial confinement fusion) ignition condition, a discussion of the triple product as a metric, a correction, and a new plot of scientific energy gain $Q_{\mathrm{sci}}$ vs.\ date achieved.

This paper is organized as follows. Section~\ref{sec:plots} provides updated and new plots, including descriptions of changes from the original paper. Section~\ref{sec:clarifications} provides clarifications and a correction to the original paper. Section~\ref{sec:new_data} discusses the new experimental results included in this paper. Section~\ref{sec:tables} provides updated and new data tables.

\section{Plots}
\label{sec:plots}
\begin{figure*}[p!]
\centerline{
\includegraphics[width=17.5cm]{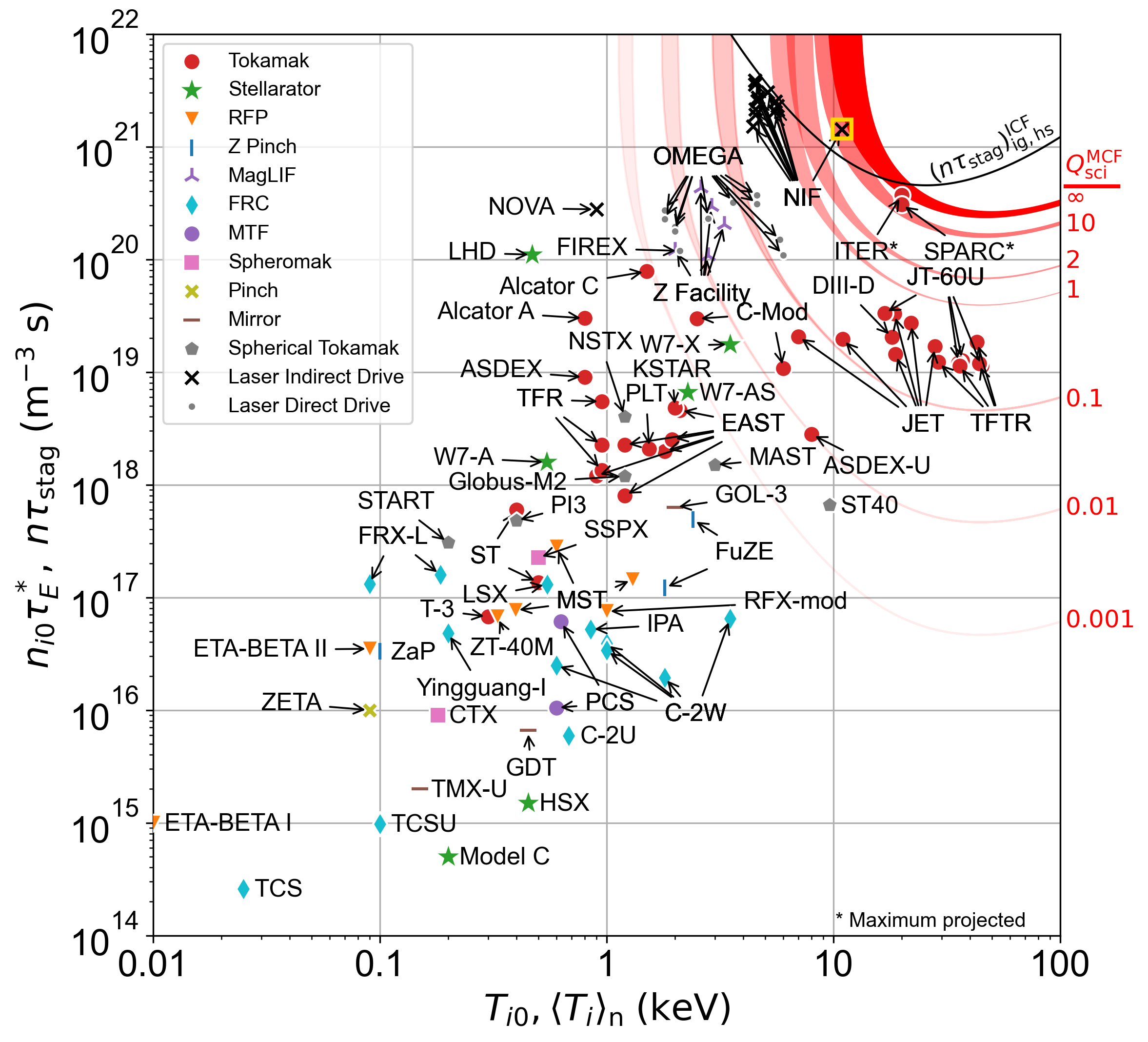}}
\caption{\label{fig:scatterplot_ntauE_vs_T} Experimentally inferred Lawson parameters ($n_{i0}\tau^*_E$ for MCF and $n\tau_{\rm stag}$ for ICF) of fusion experiments vs.\ $T_{i0}$ for MCF and $\langle T_i \rangle_{\rm n}$ for ICF (see Sec.~III of the original paper for definitions of these quantities), extracted from the published literature (see Tables~\ref{tab:mainstream_mcf_data_table}--\ref{tab:icf_mif_data_table})\@. The red contours correspond to the Lawson parameters and ion temperatures required to achieve the indicated values of scientific gain for an advanced-tokamak (AT) model, which may be considered generally representative of MCF and are labeled as $Q^{\rm MCF}_{\rm sci}$. The contour colors in the original paper have been replaced by varying shades of red here to emphasize that they originate from a tokamak model. The finite widths of the $Q_{\rm sci}^{\rm MCF}$ contours represent a range of assumed impurity levels. The black curve corresponds to the Lawson parameters and ion temperatures required to achieve hot-spot ignition and the onset of propagating burn for direct- and indirect-drive laser ICF and is labeled $(n\tau_{\rm stag})^{\rm ICF}_{\rm ig, hs}$\@. See
caption of Fig.~\ref{fig:scatterplot_nTtauE_vs_year} regarding the gold box on one of the NIF data points. For all contours, we assume representative density and temperature profiles, external-heating absorption efficiencies, and D-T fuel. For experiments that do not use D-T, the contours represent a D-T-equivalent value. See Sec.~IV of the original paper and Sec.~\ref{sec:icf_hot_spot_ignition_condition} of this paper for details on how the $Q_{\rm sci}^{\rm MCF}$ and $(n\tau_{\rm stag})^{\rm ICF}_{\rm ig, hs}$ contours are calculated and how individual data points are extracted.}
\end{figure*}

\begin{figure*}[p!]
\centerline{
\includegraphics[width=17.5cm]{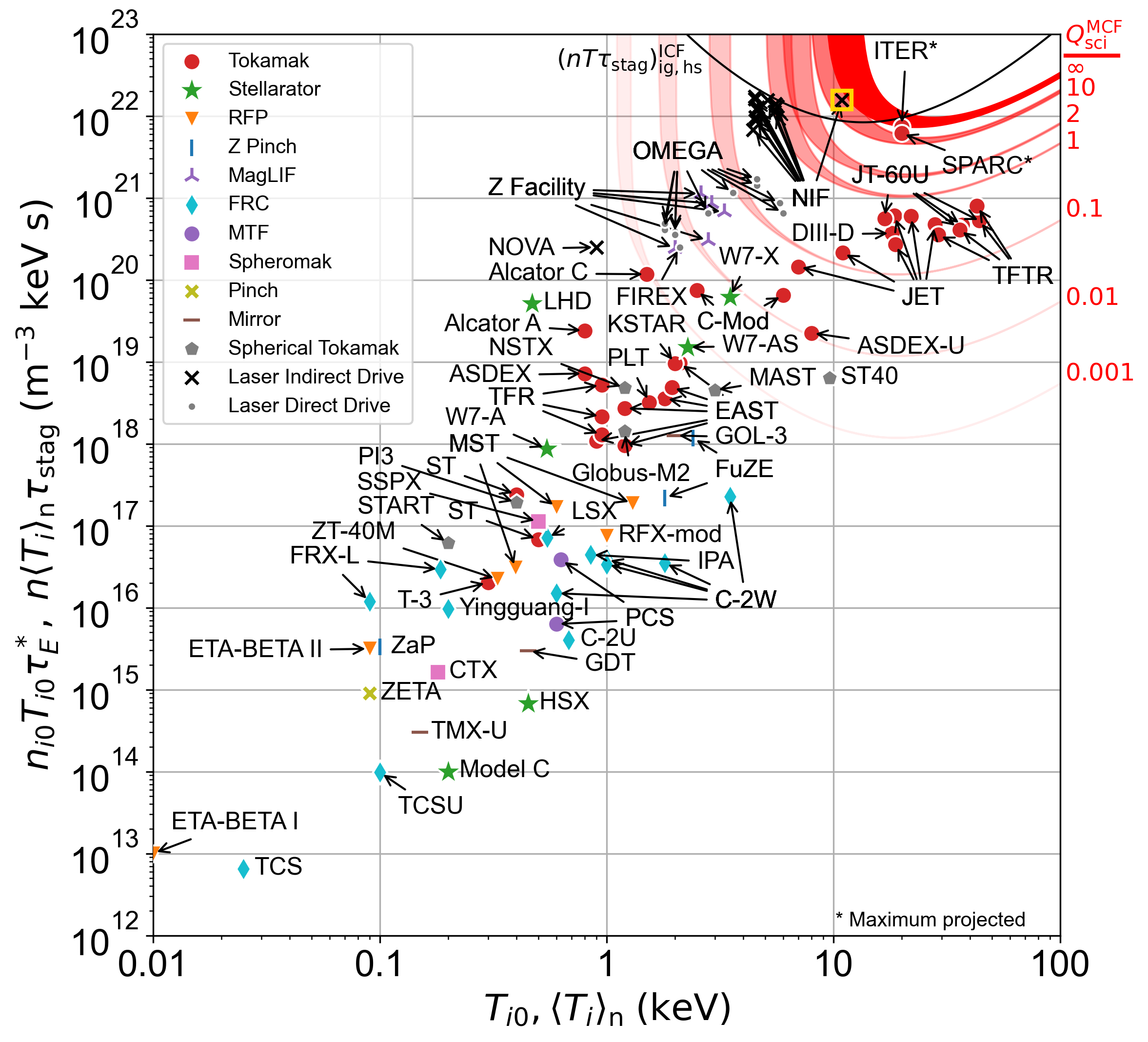}}
\caption{\label{fig:scatterplot_nTtauE_vs_T} %
Experimentally inferred triple products of fusion experiments vs. ion temperature, extracted from published literature. See the caption of Fig.~\ref{fig:scatterplot_ntauE_vs_T} for more details, and the caption of Fig.~\ref{fig:scatterplot_nTtauE_vs_year} regarding the gold box around one of the NIF data points.}
\end{figure*}

\begin{figure*}[p]
\centerline{
\includegraphics[width=17.5cm]{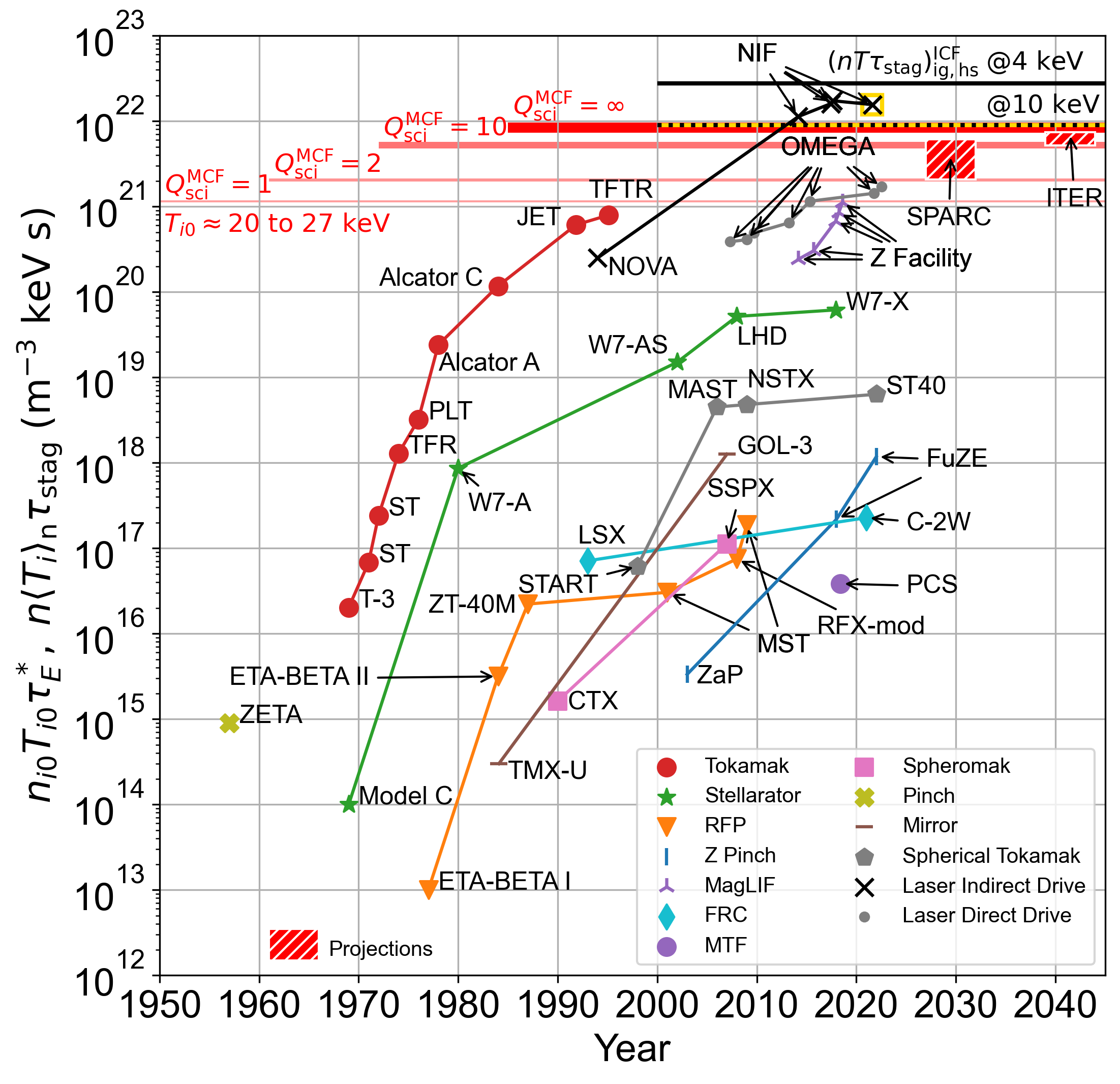}}
\caption{\label{fig:scatterplot_nTtauE_vs_year} Triple products ($n_{i0}T_{i0}\tau^*_E$ for MCF and $n \langle T_i \rangle_{\rm n} \tau_{\rm stag}$ for ICF, as defined in Sec.~III of the original paper and updated in Sec.~\ref{sec:icf_hot_spot_ignition_condition} of this paper) that set a record for a given concept vs.\ year achieved. Record values for different concepts are shown to illustrate the progress towards energy gain of different concepts over time.
The horizontal lines labeled $Q_{\rm sci}^{\rm MCF}$ represent the minimum required triple product to achieve the indicated values of $Q_{\rm sci}^{\rm MCF}$. The thickness of these lines are equal to the thickness of the equivalent contours in
Fig.~\ref{fig:scatterplot_nTtauE_vs_T} at their minimum values. The central ion temperatures required to achieve the minimum triple products (corresponding to the indicated values of scientific gain on Fig.~16 of the original paper) are in the range $20~\mathrm{keV}$ to $27~\mathrm{keV}$, which is now indicated on the plot.
The black horizontal line labeled ``$(nT\tau_{\rm stag})_{\rm ig, hs}^{\rm ICF} @ T_i = 4~\mathrm{keV}$'' represents the required triple product to achieve ignition and the onset of propagating burn in an ICF hot spot, assuming $T_i=4~\mathrm{keV}$\@.
The dashed gold-and-black horizontal line labeled ``$@ T_i = 10~\mathrm{keV}$'' represents the required triple product to achieve ignition and the onset of propagating burn in an ICF hot spot, assuming $T_i=10~\mathrm{keV}$\@.
Although the NIF shot from August 8, 2021 did not set a triple-product record, we include it in this updated plot because it achieved hot-spot ignition and the onset of propagating burn and is an appropriate terminal data point for NIF on this plot and have added a gold box to highlight this fact. This final data point for NIF lies below the black line (and was not a record triple product because stagnation time $\tau_{\rm stag}$ for an ignited fuel assembly is depressed due to the more rapid disassembly that results from the dynamics of an ignited hot spot), but it did achieve $T_i > 10~\mathrm{keV}$ and is above the dashed gold-and-black horizontal line, which denotes the requirement for hot-spot ignition and onset of propagating burn at 10~keV\@. See Sec.~\ref{sec:triple_product_limitations} of this paper and Sec.~IV B 4 of the original paper for a discussion of the limitations of using the triple product as a figure of merit for experimental results near and beyond ignition for ICF.\@
More recent and even higher-performing NIF shots are not shown on this plot but are shown on Fig. \ref{fig:Qsci_vs_year}. 
The projected triple-product ranges for SPARC and ITER are bounded above by their projected peak triple products
and below by the original stated missions of each experiment (i.e.,
$Q_{\rm fuel}^{\rm MCF}=2$ for SPARC and $Q_{\rm fuel}^{\rm MCF}=10$ for ITER)\@. The timelines for both SPARC and ITER have been delayed since the publication of the original paper, and the most recently announced timelines\cite{2024_CFS} \cite{2024_ITER} are reflected here.}
\end{figure*}

\begin{figure*}[p]
\centerline{
\includegraphics[width=17.5cm]{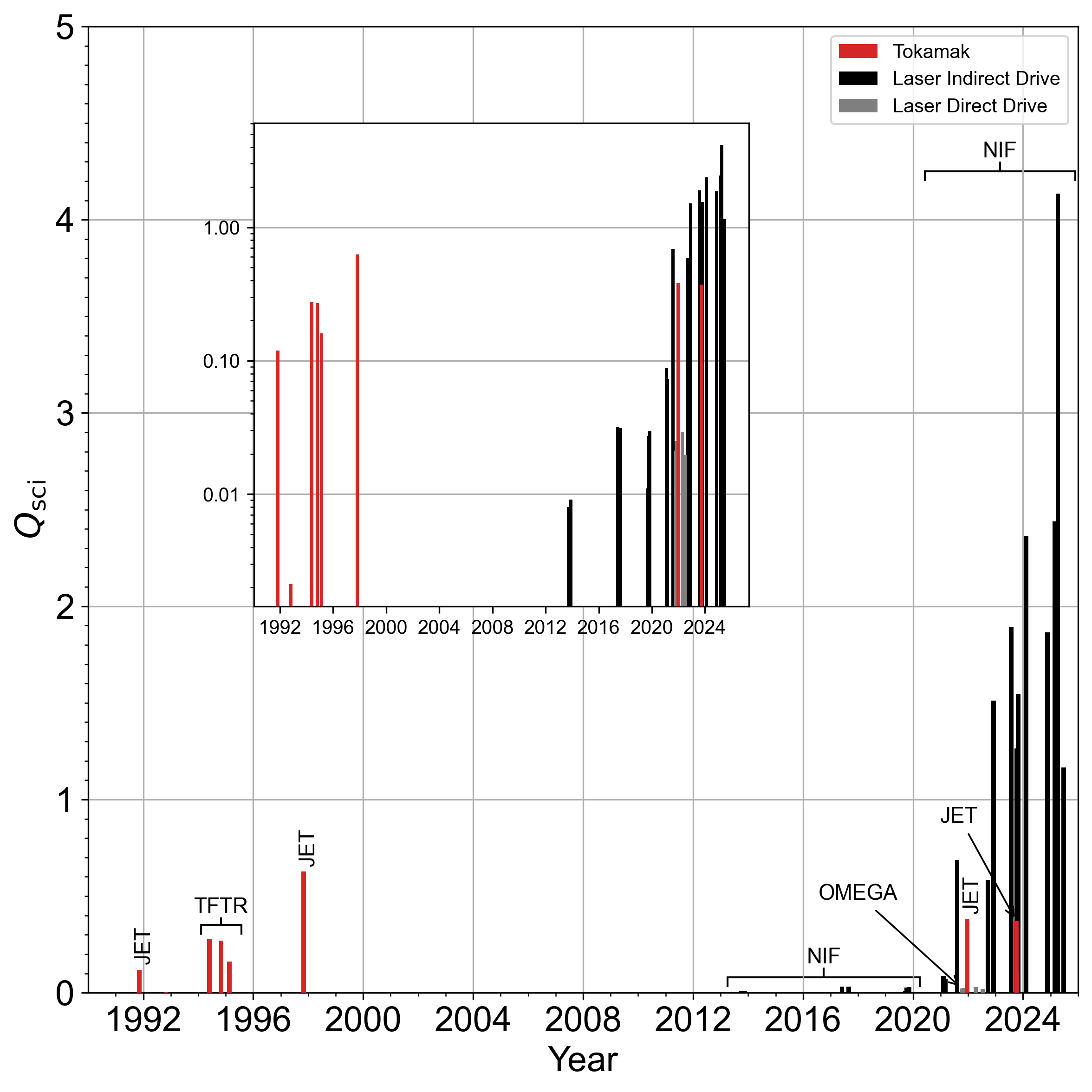}}
\caption{\label{fig:Qsci_vs_year} Plot of actual $Q_{\rm sci}$ vs.\ date achieved for D-T fusion experiments. Inset plot shows the same data on log-linear axes. Values of $Q_{\rm sci}$ for NIF and OMEGA results are reported on an energy basis ($Q_{\rm sci}=Y/E_{\rm in}$, where $Y$ is the fusion energy yield and $E_{\rm in}$ is the laser energy). Values of $Q_{\rm sci}$ for JET and TFTR are reported on a power basis ($Q_{\rm sci}=P_{\rm F}/P_{\rm in}$, where $P_{\rm F}$ is fusion power and $P_{\rm in}$ is heating power crossing the vacuum-vessel boundary). JET shots 99971 and 99972 (taken on the same day in 2021) use power levels averaged over 5 and 2~s, respectively. JET shots 104522 and 104600 (taken in 2023) use power levels averaged over 3 and 5~s, respectively. Earlier shots from JET and TFTR (from the 1990s) are instantaneous values taken at the time of peak fusion power.
\\
}
\end{figure*}

This section provides updated plots of compiled Lawson parameters vs.\ ion temperatures $T_i$ (Fig.~\ref{fig:scatterplot_ntauE_vs_T}), compiled triple products vs.\ $T_i$ (Fig.~\ref{fig:scatterplot_nTtauE_vs_T}), record triple products vs.\ date achieved (Fig.~\ref{fig:scatterplot_nTtauE_vs_year}), and a new plot of $Q_{\mathrm{sci}}$ vs.\ date achieved (Fig.~\ref{fig:Qsci_vs_year}). These plots are generated from an updated code, which is available for download.\cite{Wurzel_2022} There are seven changes from the plots in the original paper:
\begin{enumerate}

\item Additional data points from C-2W, FuZE, JET, Z Facility (referred to previously as MagLIF), NIF, OMEGA, PCS, PI3, and ST-40 are included.

\item The contour colors corresponding to different values of $Q_{\rm sci}^{\rm MCF}$ in Figs.~\ref{fig:scatterplot_ntauE_vs_T} and \ref{fig:scatterplot_nTtauE_vs_T}, as well as the corresponding lines in Fig.~\ref{fig:scatterplot_nTtauE_vs_year}, are no longer arbitrary. Instead, these contours are displayed in red, based on an advanced-tokamak (AT) model with associated temperature and density profiles and a range of impurity levels, which are described in Secs.~IV.A.2 and IV.A.3 and summarized in the caption of Fig.~22 of the original paper. The positions of the contours are unchanged. Different transparency levels are used to indicate different values of $Q_{\rm sci}^{\rm MCF}$. This change is intended to emphasize that the contours shown are specific to the AT, although they are generally representative of magnetic confinement fusion (MCF) and are therefore labeled as $Q^{\rm MCF}_{\rm sci}$.

\item The black ignition curves in Figs.~\ref{fig:scatterplot_ntauE_vs_T} and \ref{fig:scatterplot_nTtauE_vs_T}, the ignition lines in Fig.~\ref{fig:scatterplot_nTtauE_vs_year}, and data points for NIF and OMEGA (where stagnation time $\tau_{\rm stag}$ was reported) have all been increased along the $y$-axis by a factor of three. This is because the previous adjustment factor of 1/3 to $\tau_{\rm stag}$ of an ICF hot spot to obtain the idealized confinement time $\tau$ (discussed in Sec.~IV.B.2 of the original paper) is now reflected as a factor-of-three increase along the $y$-axis of the ignition curve. 
This adjustment is discussed further in Sec.~\ref{sec:icf_hot_spot_ignition_condition}.

\item The data points representing NIF shot N210808 in Figs.~\ref{fig:scatterplot_ntauE_vs_T}, ~\ref{fig:scatterplot_nTtauE_vs_T}, and \ref{fig:scatterplot_nTtauE_vs_year} include a surrounding gold box to indicate that the shot achieved hot-spot ignition and the onset of propagating burn, and is therefore the terminal data point for NIF on those figures. Additionally, a dashed black-and-gold line appears in Fig.~\ref{fig:scatterplot_nTtauE_vs_year} indicating the lower required triple product to achieve hot-spot ignition and the onset of propagating burn at $T_i=10~\mathrm{keV}$, which N210808 (but not earlier NIF shots) exceeded. These changes are discussed further in Sec.~\ref{sec:triple_product_limitations}.

\item The anticipated timelines for SPARC and ITER have been updated based on recent announcements.\cite{2024_CFS,2024_ITER}

\item Where known, Figs.~\ref{fig:scatterplot_nTtauE_vs_year} and \ref{fig:Qsci_vs_year} utilize the exact date on which the experiment occurred (only the year was used in the original paper). We welcome readers to send us the full date for any experiments showing only the year in Tables~\ref{tab:mainstream_mcf_data_table}--\ref{tab:icf_mif_data_table}.

\item In all figures, we have separated the experimental category of "Laser ICF" into ``Laser Indirect Drive'' and ``Laser Direct Drive'' to track the progress of each independently.

\end{enumerate}

In addition to the above changes, we introduce a new plot (Fig.~\ref{fig:Qsci_vs_year}) of $Q_{\rm sci}$ vs.\ date achieved. For this plot, we recognize only actual fusion energy released (for ICF) or fusion power (for MCF, see the plot caption for details).
The inclusion of this new graph is a recognition of scientific energy breakeven ($Q_{\rm sci} > 1$) having
been achieved for the first time in a controlled-fusion experiment\cite{2024_Abu-Shawareb} shortly after the publication of our original paper.
The most relevant figure of merit for such high-performing experiments is simply $Q_{\mathrm{sci}}$.
With even higher performance, a useful metric could be wall-plug gain $Q_{\rm wp}$ as defined in Eq.~(27) of the original paper.

\section{Clarifications and Corrections}
\label{sec:clarifications}

\subsection{The term ``Lawson criterion''}
In the original paper (Sec.~I, first paragraph), we stated:
\begin{displayquote}
``The required temperature and Lawson parameter for self-heating from charged fusion products to exceed all losses is known as the \textit{Lawson criterion}. A fusion plasma that has reached these conditions is said to have achieved \textit{ignition}.''
\end{displayquote}
It would have been clearer to have used \textit{Lawson criteria for ignition} instead of \textit{Lawson criterion} because there are two separate requirements (product of density $n$ and energy confinement time $\tau_E$ or pulse duration $\tau$, and, separately, $T_i$), so the plural ``criteria" is appropriate. Furthermore, although ``Lawson criteria'' (or ``Lawson criterion'') is usually assumed to mean the requirements for ignition, Lawson originally described ``criteria for a useful thermonuclear reactor,''\cite{1955Lawson} which he determined to be an energy gain of 2. Therefore, when referring to the {\em Lawson criteria}, it is worthwhile to specify the criteria being discussed, e.g., the criteria for ignition or the criteria for a specific value of energy gain.


\subsection{Necessity of ignition}

In the original paper (Sec.~I, first paragraph), we stated:
\begin{displayquote}
``Although ignition is not required for a commercial fusion-energy system, \ldots''
\end{displayquote}
It would have been more accurate to state ``Although ignition is not required for
a commercial fusion-energy system for certain fusion approaches, \ldots'' While MCF and MIF concepts may not strictly need to reach ignition for a commercially viable system  (and ignition may in fact not be desirable from the perspective of plasma control), ICF concepts likely need to achieve
ignition to realize the energy gains needed for a commercially viable system.

\subsection{Reframing the ICF ignition condition using experimentally measured $\tau_{\rm stag}$}
\label{sec:icf_hot_spot_ignition_condition}

In Sec.~IV.B.2 of the original paper, we described reducing experimentally measured values of $\tau_{\rm stag}$, which is defined as the full width at half maximum (FWHM) of the burn duration, by a factor of three. 
This adjustment was based on the requirement for ICF hot-spot ignition and the onset of propagating burn (herein referred to as simply ``ignition" for both direct- and indirect-drive), as described by Christopherson et al.~\cite{Christopherson_2019}

In this update,
instead of reducing the measured $\tau_{\rm stag}$ by a factor of three, we increase the ignition threshold of the Lawson parameter (and triple product) from the idealized case by the same factor of three. 
As a result, all ICF data points with reported $\tau_{\rm stag}$ (except FIREX) and
the ignition contour itself are shifted upward along the $y$-axis by a factor of three.

The revised ignition contour is now labeled $(n\tau_{\rm stag})^{\rm ICF}_{\rm ig, hs}$ to emphasize that the criterion is based on the experimentally measurable stagnation time, $\tau_{\rm stag}$, rather than the idealized stagnation duration $\tau$. This updated formulation aligns conceptually with our approach to MCF, where $Q_{\rm sci}$ contours are adjusted to account for deviations from the idealized case (i.e., flat profiles and no impurities). 
The physical ICF ignition condition can be derived from the ``Christopherson criterion'' $f_\alpha \geq 1.4$, where
\begin{equation}
    f_{\alpha}=\frac{1}{2}\frac{\theta_\alpha E_\alpha}{E_{\rm hs}}.
\end{equation}
Here, $\theta_\alpha$ is the fraction of alpha-particle energy deposited in the hot spot, $E_\alpha$ is the total alpha energy from fusion reactions, and $E_{\rm hs}$ is the internal energy of the hot spot at bang time (time of maximum neutron production). Physically, $f_\alpha$ is the ratio of alpha-particle energy deposited into the hot spot to hot-spot internal energy at bang time. Because approximately half of the alphas have deposited their energy by bang time, a factor of 1/2 is included.

In Sec.~III.F of the original paper, the idealized ignition condition was derived by balancing the total alpha-particle energy with the internal energy of an idealized hot spot (which assumes an instantaneous temperature rise sustained over a duration $\tau$). 
Applying the physical ignition condition above leads to the ignition threshold in terms of $\tau_{\rm stag}$:
\begin{equation}
    \label{eq:ignition_threshold}
    (n \tau_{\rm stag})^{\rm ICF}_{\rm ig, hs} = \frac{2 f_{\alpha}}{\theta_\alpha}\frac{12T}{\langle \sigma v \rangle \epsilon_\alpha}.
\end{equation}

Substituting $f_\alpha = 1.4$ and $\theta_\alpha = 0.93$ (for NIF conditions\cite{Christopherson_2019}) results in a threefold increase in the ignition threshold relative to the idealized case. Although this is intended to be a requirement for both direct- and indirect-drive hot-spot ignition and onset of propagating burn, the factor-of-three increase from the ideal case would vary if $\theta_\alpha$ in Eq.~\ref{eq:ignition_threshold} differs from 0.93.


\subsection{Limitations of the triple product as a metric}
\label{sec:triple_product_limitations}
The triple product is a convenient figure of merit for evaluating progress \textit{toward} fusion breakeven and gain because it is a single scalar value that can be compared across experiments and tracked over time, as in Fig. \ref{fig:scatterplot_nTtauE_vs_year}. Mathematically, it is the area of a rectangle defined by the origin and any point within the domain of Fig. \ref{fig:scatterplot_ntauE_vs_T}
(assuming linear rather than logarithmic axes).
The minimum triple product for a particular contour of $Q_{\rm sci}$ would correspond to the minimum-area rectangle with a vertex on the contour.

As emphasized in the original paper, the triple product does \textit{not} uniquely map to a value of $Q_{\rm sci}$ (in the case of MCF) or to the threshold of hot-spot ignition and onset of propagating burn (in the case of ICF)\@. However, specific combinations of $T_i$ and $n\tau_E^*$ (in the case of MCF) or $n\tau_{\rm stag}$ (in the case of ICF) do map to a unique value of $Q_{\rm sci}$ or the ignition threshold, respectively. For this reason, Fig.~\ref{fig:scatterplot_nTtauE_vs_year} indicates the assumed $T_i$ for the various horizontal lines. In general, only density and (energy) confinement time may be traded off with each other without altering the achieved gain. {\it Temperature cannot be traded off with the other two quantities and should be carefully scrutinized when evaluating any triple-product claim.}

In the context of ICF, further improvements in energy gain beyond ignition are driven by details such as the degree of propagating burn and burn-up fraction of the surrounding fuel layer that are not directly encapsulated by the triple product. Therefore, beyond ignition, the triple product is no longer a useful metric to quantify improvements in ICF performance. In fact, the triple product for the highest-performing NIF shots (based on $Q_{\mathrm{sci}}$) declines due to propagating burn, which causes more rapid disassembly and therefore lower values of $\tau_{\rm stag}$. For these reasons, shot N210808 is the terminal data point for NIF in Figs.~\ref{fig:scatterplot_ntauE_vs_T}, \ref{fig:scatterplot_nTtauE_vs_T}, and \ref{fig:scatterplot_nTtauE_vs_year}.

In the context of MCF, improvements in triple product beyond scientific energy breakeven should correlate with increased energy gain if the temperature is in the range for which this occurs (approximately 10--30~keV for deuterium-tritium; see Fig.~\ref{fig:scatterplot_nTtauE_vs_T}). A limitation of the triple product as a figure of merit in the MCF context is illustrated in Fig.~\ref{fig:scatterplot_nTtauE_vs_year}, where TFTR holds the triple-product record for tokamaks due to its exceptionally high ion temperatures ($>40~\mathrm{keV}$) achieved in 1996. In contrast, JET achieved the highest $Q_{\rm sci}$ in a tokamak in 1997 (see Fig.~\ref{fig:Qsci_vs_year}). However, this shot was not a triple-product record for tokamaks and therefore does not appear in Fig.~\ref{fig:scatterplot_nTtauE_vs_year}. Additional considerations when evaluating a reported triple product and its mapping to a value of $Q_{\rm sci}$ are discussed in Sec.~IV of the original paper.

\subsection{Correction}
Equation~(23) of the the original paper had ``$\langle \sigma v \rangle(T)$'' in the denominator, which was intended to indicate that $\langle \sigma v \rangle$ is a function of $T$. Because this introduces ambiguity
and because the rest of the original paper does not explicitly show the $T$ dependence of $\langle \sigma v \rangle$, Eq.~(23) of the original paper should be corrected:
\begin{equation}
\label{eq:corrected_triple_product_steady_state}
    nT\tau_E = \frac{3T^{2}}{(f_c+ Q_{\rm fuel}^{-1})\langle \sigma v  \rangle \epsilon_{F}/4 - C_{B}T^{1/2}}.
\end{equation}

\section{New data}
\label{sec:new_data}
Since the original paper was published in 2022, new experimental results have been published. Notes on each new data point are included in this section.

\subsection{NIF}
\label{sec:NIF_results}
On August 8th, 2021 the National Ignition Facility (NIF) achieved hot-spot ignition and the onset of propagating burn, but not ignition according to the definition\cite{NAS_1997} of the National Academies of Sciences, Engineering, and Medicine (NASEM), the details of which are discussed below. When we submitted our original paper for publication in December~2021, the only publication describing this experimental shot (N210808) was an initial technical report,\cite{Hurricane_2022} from which we extracted relevant parameters. Since then, peer-reviewed publications\cite{2022_Kritcher,2024_Abu-Shawareb} have described this shot in detail and included minor adjustments to the reported measurements that are used in this paper.

On December 5th, 2022, ignition according to the NASEM definition was achieved in shot N221204.\cite{2024_Abu-Shawareb} NASEM defined ignition as the ``ratio of fusion yield to laser energy'' exceeding unity,\cite{NAS_1997} which corresponds to our definition of scientific energy breakeven. Shot N221204 crossed this threshold, achieving $Q_{\rm sci}=1.5$. Since then, eight additional shots have achieved the NASEM definition of ignition with a record of $Q_{\rm sci} = 4.13$ achieved in April 2025.\cite{2025_LLNL} As of this writing, details of the eight additional shots have not yet been published in a peer-reviewed journal, but they are included in Fig.~\ref{fig:Qsci_vs_year} and Table~\ref{tab:q_sci_data_table} with the exception of the most recent shot from June 2025 for which the laser energy has not yet been reported.

We also include minor updates to the reported, measured parameters achieved in shots N210207, N210307,  N210808, as well as newer shots N220919 and N221204, based on the most recent peer-reviewed publication providing these data.\cite{2024_Abu-Shawareb}

Based on the discussion of Sec.~\ref{sec:triple_product_limitations}, N210808 is the final NIF result included in Table~\ref{tab:icf_mif_data_table} and in Figs.~\ref{fig:scatterplot_ntauE_vs_T}, \ref{fig:scatterplot_nTtauE_vs_T}, and \ref{fig:scatterplot_nTtauE_vs_year}, with subsequent results included in Table~\ref{tab:q_sci_data_table} and Fig.~\ref{fig:Qsci_vs_year}.

\subsection{OMEGA}
Details of OMEGA's 2021 experimental campaign were published\cite{2024_Williams} in early 2024 showing improved performance and the first achievement of fuel energy gain (ratio of fusion energy to the hot-spot internal energy greater than unity) in a direct-drive laser-ICF experiment. This result is the highest achieved $Q_{\rm sci}$ in a direct-drive laser-ICF experiment that has been published, having achieved $Q_{\mathrm{sci}}=0.02$, and is included in Table~\ref{tab:q_sci_data_table} and Fig.~\ref{fig:Qsci_vs_year}.

A separate publication\cite{2024_Gopalaswamy} reported results for shot 104949, which, if scaled up in size, would be projected to achieve a burning plasma, but the shot did not (and was not intended to) set a record for fusion-energy yield. Due to the high achieved pressure, it set a laser direct-drive triple-product record, which is reflected in Table~\ref{tab:icf_mif_data_table} and Fig.~\ref{fig:scatterplot_nTtauE_vs_year}. The energy yield from this shot was extrapolated from the number of reported D-T neutrons (energy from other fusion reactions are neglected) and is reflected in Table~\ref{tab:q_sci_data_table} and Fig.~\ref{fig:Qsci_vs_year}.

\subsection{Z Facility}
A new publication\cite{2022_Knapp} reports record results for the MagLIF platform at the Z Facility\@. Shot z3289 achieved a high triple product, exceeding $10^{21}~\mathrm{keV\,m^{-3}\,s}$. Note that although it appears at the red $Q_\mathrm{sci}^\mathrm{MCF}=1$ line in Fig.~\ref{fig:scatterplot_nTtauE_vs_year}, it does not correspond to that line because $T_i$ was too low ($2.6~\mathrm{keV}$), and the red lines assume a temperature of approximately $20~\mathrm{keV}$ (and advanced-tokamak profiles). Recent work\cite{2025_Alexander} investigates ignition conditions relevant to the MagLIF concept.

Fig.~\ref{fig:scatterplot_ntauE_vs_T} more clearly shows the relative performance of the Z Facility results and the benefit that higher temperature would have in advancing performance. Because the MagLIF platform on the Z Facility utilizes pure deuterium fuel, their results are not included in Table~\ref{tab:q_sci_data_table} nor in Fig.~\ref{fig:Qsci_vs_year}.

\subsection{JET}
Since publication of the original paper, JET has published the results of two recent deuterium-tritium experiment (DTE) campaigns, DTE2 and DTE3, which occurred in 2021 and 2023, respectively. The goals of these campaigns included demonstrating long-pulse operation and the development of experimental scenarios for ITER. These campaigns were not designed to set scientific gain or triple-product records.
They did, however, achieve record levels of fusion energy produced in a single fusion experiment, with shot 99971 producing 59~MJ of fusion energy during the DTE2 campaign and shot 104522 producing 69~MJ of fusion energy during the DTE3 campaign. The scientific energy gains achieved by these shots were 0.33 and 0.37, respectively. Published temperature, density, and energy-confinement-time data from shot 99972 (from DTE2) and shot 104600 (from DTE3) were extracted and are included in Figs.~\ref{fig:scatterplot_ntauE_vs_T} and \ref{fig:scatterplot_nTtauE_vs_T}.

\subsection{ST40}
Tokamak Energy published results from ST40\cite{2023_McNamara} demonstrating a central 
$T_i$ of $9.6~\mathrm{keV}$ and a central ion density $n_i$ of $5.5 \times 10^{19}~\mathrm{m}^{-3}$ at a flat-top time of $56~\mathrm{ms}$ into the pulse. To evaluate $\tau_E^*$, we take the stored thermal energy to be 70\% of $35~\mathrm{kJ}$ because the publication states “fast ions account for approximately 30\% of the stored energy,” and their Fig.~1 indicates a stored energy of $35~\mathrm{kJ}$ at $56~\mathrm{ms}$\@. The absorbed heating power includes 92\% of $1.6~\mathrm{MW}$ of neutral-beam heating (accounting for 8\% maximum reported shine-through) and $0.5~\mathrm{MW}$ of ohmic heating for a total absorbed heating power of approximately $2~\mathrm{MW}$\@. The $\tau_E^*$ is therefore $12~\mathrm{ms}$, and the triple product is evaluated to be $6.3 \times 10^{18}~\mathrm{keV\,m^{-3}\,s}$, which is in agreement with that reported by Tokamak Energy.

\subsection{FuZE}
New experimental results from the FuZE experiment\cite{2024_Goyon} at Zap Energy indicate record performance of their stabilized Z-pinch.
In shots at a plenum pressure of 200 psi, electron temperatures of $2.4~\mathrm{keV}$ and electron densities of $4.9\times 10^{23}~\mathrm{m}^{-3}$ were reported. The energy confinement time is assumed to be 1~$\mu \mathrm{s}$ based on the flow-through time of the plasma along the pinch as discussed in the original paper. This assumption (that advection along the pinch is the dominant heat-loss mechanism) would benefit from further study.

\subsection{C-2W}
A new publication from TAE Technologies \cite{2024_Gota} provides data demonstrating
electron temperature $T_e \approx 1$~keV (to our knowledge, for the first time ever) in a field-reversed configuration (FRC) (shot~122588 on C-2W)\@. Although the density of this shot was not published, the density was in the range $1.2$--$1.4 \times 10^{19}~\mathrm{m^{-3}}$.\cite{2024_Gota_Private} The $\tau_E^{*}$ for shot~122588 is inferred to be $2~\mathrm{ms}$ (from the slope of Fig.~4 of the aforementioned publication). Because $T_i$ for previously published shots were higher ($\approx 3.5~\mathrm{keV}$), shot~122588 is not a triple-product record and therefore does not appear in Fig.~\ref{fig:scatterplot_nTtauE_vs_year}.

\subsection{PCS}
A new publication\cite{2024_Howard} analyzes the results of General Fusion's Plasma Compression Science (PCS) program. PCS was a campaign of magnetized-target-fusion (MTF) experiments in which a compact toroid was formed by a Marshall gun and injected into a spheroidal chamber with a current-carrying center post. The combined poloidal (due to magnetization during formation) and toroidal (due to the current-carrying center post) magnetic fields form a spherical tokamak within the chamber. At this point, a high-explosive charge is detonated to implode the chamber (acting like a liner) and compress the plasma.

Key parameters for this concept (and all fusion concepts that involve compressional heating) are the particle confinement time $\tau_p$, $\tau_E$, and the time taken to compress the plasma $\tau_C$. For density to increase during compression, $\tau_C \ll \tau_p$ is required. For the temperature to increase, the compression must occur quasi-adiabatically, i.e., $\tau_C \ll \tau_E$ is required.

In the best-performing experiment, PCS-16, $\tau_p$ was much larger than $\tau_C= 139~\mathrm{\mu s}$, and the electron density $n_e$ increased from $1.2\times 10^{20}~\mathrm{m^{-3}}$ before compression to $2.3 \times 10^{22}~\mathrm{m^{-3}}$ at peak compression. 
However, because $\tau_{E_{e}} \approx 50~\mathrm{\mu s}$ was less than half of $\tau_C= 139~\mathrm{\mu s}$, the pre-compression $T_e$ of 200~eV did not increase, and the peak $T_i$ only reached 629~eV from a pre-compression value of 600~eV\@.

The liner begins to move at $t=300~\mathrm{\mu s}$, at which point the plasma is uncompressed. At this time $n_e$ is $1.25 \times 10^{20}~\mathrm{m^{-3}}$, $T_e$ is $207~\mathrm{eV}$, and $T_i$ is $600~\mathrm{eV}$\@.
The pre-compression $\tau_E$ is reported as $84~\mathrm{\mu s}$ based on MHD simulations that matched experimental data on the decay of measured magnetic fields. The pre-compression triple product is therefore $6.3 \times 10^{15}~\mathrm{keV\, m^{-3}\,s }$.

At $t=405~\mathrm{\mu s}$, the compressed plasma begins to lose MHD stability. We evaluate the parameters at this point of the compression sequence when the average $n_e$ is $5\times10^{20}~{\mathrm m^{-3}}$, the $T_e$ is approximately $180~\mathrm{eV}$, and the $T_i$ is $627~\mathrm{eV}$. The energy confinement time of ions at this point $\tau_{\mathrm Ei}$ is reported as $122~\mathrm{\mu s}$. We use this value as the effective $\tau_E$ because it encapsulates all cooling channels for ions and is shorter than the electron-ion equilibration time $\tau_{\mathrm eq}$ of $693~{\mathrm \mu s}$. This implies that the dominant cooling mechanism of ions in this particular
case is not heat transfer to electrons.

We caution that the effective confinement time $\tau_{\rm eff}$~\cite{2022_Wurzel_Hsu} could be reduced  from $\tau_E$ by half if $\tau_{\rm stag}$ (also not reported) is of the same order as $\tau_E$ at peak compression. In this case, the effective triple product would be reduced by the same factor.


\subsection{PI3}
General Fusion recently published results\cite{2025_Tancetti} for the upgraded PI3 spherical tokamak, which began operation in 2021. Although intended to study spherical tokamaks to understand the timescales needed for compression in future machines, PI3 experiments do not perform compression. Shot 21 100
reported a $T_i$ of $400~\mathrm{eV}$, a central $n_e$ of $4\times10^{19}~\mathrm{m^{-3}}$, and a
$\tau_E$ of approximately $12~\mathrm{ms}$.

\section{Tables}
\label{sec:tables}
Tables~\ref{tab:mainstream_mcf_data_table}--\ref{tab:icf_mif_data_table}, which provide
the data for Figs.~\ref{fig:scatterplot_ntauE_vs_T}--\ref{fig:scatterplot_nTtauE_vs_year}, are updated from
their corresponding tables in the original paper and include new data points and 
new references upon which the new data points are based. Data on stellarators has been shifted from Table~\ref{tab:mainstream_mcf_data_table} to Table~\ref{tab:alternates_mcf_data_table} to keep each table to a single page. 
Table~\ref{tab:q_sci_data_table} provides the numerical data for Fig.~\ref{fig:Qsci_vs_year}. For all tables, data unchanged from the original paper cite the original paper (which cite{\bf s} the primary source). Updated or new data cite the primary source.


\begin{sidewaystable*}[p]
\caption{Data for tokamaks and spherical tokamaks.}
\label{tab:mainstream_mcf_data_table}
\begin{tabular*}{\textwidth}{@{\extracolsep{\fill}}llllllllllll}
\hline\noalign{\smallskip}
Project & Concept & Date & Shot identifier & Reference & \thead{$T_{i0}$ \\ (\si{keV})} & \thead{$T_{e0}$ \\ (\si{keV})} & \thead{$n_{i0}$ \\ (\si{m^{-3}})} & \thead{$n_{e0}$ \\ (\si{m^{-3}})} & \thead{$\tau_{E}^{*}$ \\ (\si{s})} & \thead{$n_{i0} \tau_{E}^{*}$ \\ (\si{m^{-3}~s})} & \thead{$n_{i0} T_{i0} \tau_{E}^{*}$ \\ (\si{keV~m^{-3}~s})} \\
\noalign{\smallskip}\hline\noalign{\smallskip}
T-3 & Tokamak & 1969 & $H_z=25$kOe, $I_z=85$kA discharges & \onlinecite{2022_Wurzel_Hsu} & \num{0.3} & \num{1.05} & \num{2.25e+19}$^{\ddagger}$ & \num{2.25e+19} & \num{0.003} & \num{6.8e+16} & \num{2.0e+16} \\
ST & Tokamak & 1971 & 10cm limiter, 42 kA & \onlinecite{2022_Wurzel_Hsu} & \num{0.5} & \num{1.45} & \num{4e+19}$^{\ddagger}$ & \num{4e+19} & \num{0.0034} & \num{1.4e+17} & \num{6.8e+16} \\
ST & Tokamak & 1972 & 12cm limiter & \onlinecite{2022_Wurzel_Hsu} & \num{0.4} & \num{0.8} & \num{6e+19}$^{\ddagger}$ & \num{6e+19} & \num{0.01} & \num{6.0e+17} & \num{2.4e+17} \\
TFR & Tokamak & 1974 & Molybdenum limiter & \onlinecite{2022_Wurzel_Hsu} & \num{0.95} & \num{1.8} & \num{7.1e+19}$^{\ddagger}$ & \num{7.1e+19} & \num{0.019} & \num{1.3e+18} & \num{1.3e+18} \\
PLT & Tokamak & 1976 & 22149-231 & \onlinecite{2022_Wurzel_Hsu} & \num{1.54} & \num{1.86} & \num{5.2e+19}$^{\ddagger}$ & \num{5.2e+19} & \num{0.04} & \num{2.1e+18} & \num{3.2e+18} \\
Alcator A & Tokamak & 1978 & 8.7T discharge & \onlinecite{2022_Wurzel_Hsu} & \num{0.8} & \num{0.9} & \num{1.5e+21}$^{\ddagger}$ & \num{1.5e+21} & \num{0.02} & \num{3.0e+19} & \num{2.4e+19} \\
TFR & Tokamak & 1981 & Iconel limiter & \onlinecite{2022_Wurzel_Hsu} & \num{0.95} & \num{1.2} & \num{1.61e+20}$^{\ddagger}$ & \num{1.61e+20} & \num{0.034} & \num{5.5e+18} & \num{5.2e+18} \\
TFR & Tokamak & 1982 & Carbon limiter & \onlinecite{2022_Wurzel_Hsu} & \num{0.95} & \num{1.5} & \num{9e+19}$^{\ddagger}$ & \num{9e+19} & \num{0.025} & \num{2.2e+18} & \num{2.1e+18} \\
Alcator C & Tokamak & 1984 & Multiple pellet injection & \onlinecite{2022_Wurzel_Hsu} & \num{1.5} & \num{1.5} & \num{1.5e+21}$^{\ddagger}$ & \num{1.5e+21} & \num{0.052} & \num{7.8e+19} & \num{1.2e+20} \\
ASDEX & Tokamak & 1988 & 23349-57 & \onlinecite{2022_Wurzel_Hsu} & \num{0.8} & \num{1} & \num{7.50e+19}$^{\ddagger *}$ & – & \num{0.12} & \num{9.0e+18} & \num{7.2e+18} \\
JET & Tokamak & 1991-11-02 & 26087 & \onlinecite{2022_Wurzel_Hsu} & \num{18.6} & \num{10.5} & \num{4.1e+19} & \num{5.1e+19} & \num{0.8}$^{\#}$ & \num{3.3e+19} & \num{6.1e+20} \\
JET & Tokamak & 1991-11-04 & 26095 & \onlinecite{2022_Wurzel_Hsu} & \num{22.0} & \num{11.9} & \num{3.4e+19} & \num{4.5e+19} & \num{0.8}$^{\#}$ & \num{2.7e+19} & \num{6.0e+20} \\
JET & Tokamak & 1991-11-09 & 26148 & \onlinecite{2022_Wurzel_Hsu} & \num{18.8} & \num{9.9} & \num{2.4e+19} & \num{3.6e+19} & \num{0.6}$^{\#}$ & \num{1.4e+19} & \num{2.7e+20} \\
TFTR & Tokamak & 1992-10-29 & 68522 & \onlinecite{2022_Wurzel_Hsu} & \num{29.0} & \num{11.7} & \num{6.8e+19} & \num{9.6e+19} & \num{0.18}$^{\#}$ & \num{1.2e+19} & \num{3.5e+20} \\
JT-60U & Tokamak & 1994 & 17110 & \onlinecite{2022_Wurzel_Hsu} & \num{37} & \num{12} & \num{4.2e+19} & \num{5.5e+19} & \num{0.3}$^{\#}$ & \num{1.3e+19} & \num{4.7e+20} \\
TFTR & Tokamak & 1994-05-27 & 76778 & \onlinecite{2022_Wurzel_Hsu} & \num{44} & \num{11.5} & \num{6.3e+19} & \num{8.5e+19} & \num{0.19}$^{\#}$ & \num{1.2e+19} & \num{5.3e+20} \\
TFTR & Tokamak & 1994-11-02 & 80539 & \onlinecite{2022_Wurzel_Hsu} & \num{36} & \num{13} & \num{6.7e+19} & \num{1.02e+20} & \num{0.17}$^{\#}$ & \num{1.1e+19} & \num{4.1e+20} \\
TFTR & Tokamak & 1995-02-17 & 83546 & \onlinecite{2022_Wurzel_Hsu} & \num{43} & \num{12.0} & \num{6.6e+19} & \num{8.5e+19} & \num{0.28}$^{\#}$ & \num{1.8e+19} & \num{7.9e+20} \\
JT-60U & Tokamak & 1996 & E26939 & \onlinecite{2022_Wurzel_Hsu} & \num{45.0} & \num{10.6} & \num{4.35e+19} & \num{6e+19} & \num{0.26}$^{\#}$ & \num{1.1e+19} & \num{5.1e+20} \\
JT-60U & Tokamak & 1996 & E26949 & \onlinecite{2022_Wurzel_Hsu} & \num{35.5} & \num{11.0} & \num{4.3e+19} & \num{5.85e+19} & \num{0.28}$^{\#}$ & \num{1.2e+19} & \num{4.3e+20} \\
DIII-D & Tokamak & 1997 & 87977 & \onlinecite{2022_Wurzel_Hsu} & \num{18.1} & \num{7.5} & \num{8.5e+19} & \num{1e+20} & \num{0.24}$^{\#}$ & \num{2.0e+19} & \num{3.7e+20} \\
JET & Tokamak & 1997-10-31 & 42976 & \onlinecite{2022_Wurzel_Hsu} & \num{28} & \num{14} & \num{3.3e+19} & \num{4.1e+19} & \num{0.51}$^{\#}$ & \num{1.7e+19} & \num{4.7e+20} \\
JT-60U & Tokamak & 1998 & E31872 & \onlinecite{2022_Wurzel_Hsu} & \num{16.8} & \num{7.2} & \num{4.8e+19} & \num{8.5e+19} & \num{0.69}$^{\#}$ & \num{3.3e+19} & \num{5.6e+20} \\
START & Spherical Tokamak & 1998 & 35533 & \onlinecite{2022_Wurzel_Hsu} & \num{0.2}$^{\dagger}$ & \num{0.2} & \num{1.02e+20}$^{\ddagger *}$ & – & \num{0.003} & \num{3.1e+17} & \num{6.1e+16} \\
MAST & Spherical Tokamak & 2006 & 14626 & \onlinecite{2022_Wurzel_Hsu} & \num{3} & \num{2} & \num{3e+19}$^{\ddagger}$ & \num{3e+19} & \num{0.05} & \num{1.5e+18} & \num{4.5e+18} \\
NSTX & Spherical Tokamak & 2009 & 129041 & \onlinecite{2022_Wurzel_Hsu} & \num{1.2} & \num{1.2} & \num{5e+19}$^{\ddagger}$ & \num{5e+19} & \num{0.08} & \num{4.0e+18} & \num{4.8e+18} \\
EAST & Tokamak & 2012 & 41195 & \onlinecite{2022_Wurzel_Hsu} & \num{0.9} & – & \num{3e+19}$^{\ddagger}$ & \num{3e+19} & \num{0.04} & \num{1.2e+18} & \num{1.1e+18} \\
EAST & Tokamak & 2012 & 41079 & \onlinecite{2022_Wurzel_Hsu} & \num{1.2}$^{\dagger}$ & \num{1.2} & \num{2e+19}$^{\ddagger}$ & \num{2e+19} & \num{0.04} & \num{8.0e+17} & \num{9.6e+17} \\
EAST & Tokamak & 2014 & 48068 & \onlinecite{2022_Wurzel_Hsu} & \num{1.2} & – & \num{6.1e+19}$^{\ddagger}$ & \num{6.1e+19} & \num{0.037} & \num{2.3e+18} & \num{2.7e+18} \\
KSTAR & Tokamak & 2014 & 7081 & \onlinecite{2022_Wurzel_Hsu} & \num{2} & – & \num{4.80e+19}$^{\ddagger *}$ & – & \num{0.1} & \num{4.8e+18} & \num{9.6e+18} \\
EAST & Tokamak & 2015 & 56933 & \onlinecite{2022_Wurzel_Hsu} & \num{2.1} & \num{1.8} & \num{8.5e+19}$^{\ddagger}$ & \num{8.5e+19} & \num{0.054} & \num{4.6e+18} & \num{9.6e+18} \\
C-Mod & Tokamak & 2016 & 1160930042 & \onlinecite{2022_Wurzel_Hsu} & \num{6}$^{\dagger}$ & \num{6} & \num{2e+20}$^{\ddagger}$ & \num{2e+20} & \num{0.054} & \num{1.1e+19} & \num{6.5e+19} \\
EAST & Tokamak & 2016 & 71320 & \onlinecite{2022_Wurzel_Hsu} & \num{1.8} & \num{2.0} & \num{5.5e+19}$^{\ddagger}$ & \num{5.5e+19} & \num{0.036} & \num{2.0e+18} & \num{3.6e+18} \\
C-Mod & Tokamak & 2016 & 1160930033 & \onlinecite{2022_Wurzel_Hsu} & \num{2.5}$^{\dagger}$ & \num{2.5} & \num{5.5e+20}$^{\ddagger}$ & \num{5.5e+20} & \num{0.054} & \num{3.0e+19} & \num{7.4e+19} \\
ASDEX-U & Tokamak & 2016 & 32305 & \onlinecite{2022_Wurzel_Hsu} & \num{8} & \num{5} & \num{5e+19}$^{\ddagger}$ & \num{5e+19} & \num{0.056} & \num{2.8e+18} & \num{2.2e+19} \\
EAST & Tokamak & 2018 & 78723 & \onlinecite{2022_Wurzel_Hsu} & \num{1.94} & – & \num{5.58e+19}$^{\ddagger}$ & \num{5.58e+19} & \num{0.045} & \num{2.5e+18} & \num{4.9e+18} \\
Globus-M2 & Spherical Tokamak & 2019 & 37873 & \onlinecite{2022_Wurzel_Hsu} & \num{1.2} & – & \num{1.19e+20}$^{\ddagger *}$ & – & \num{0.01} & \num{1.2e+18} & \num{1.4e+18} \\
PI3 & Spherical Tokamak & 2021 & 21 100 & \onlinecite{2025_Tancetti} & \num{0.4} & \num{0.350} & \num{4e+19}$^{\ddagger}$ & \num{4e+19} & \num{0.012} & \num{4.8e+17} & \num{1.9e+17} \\
JET & Tokamak & 2021-12-21 & 99972 & \onlinecite{2023_Maslov} & \num{11} & – & \num{8.5e+19}$^{\ddagger}$ & \num{8.5e+19} & \num{0.23} & \num{2.0e+19} & \num{2.2e+20} \\
ST40 & Spherical Tokamak & 2022 & 10009 & \onlinecite{2023_McNamara} & \num{9.6} & \num{3} & \num{5.5e+19} & \num{9e+19} & \num{0.012} & \num{6.6e+17} & \num{6.3e+18} \\
JET & Tokamak & 2023-10-09 & 104600 & \onlinecite{2025_Kappatou,2024_Carvalho} & \num{7}$^{\dagger}$ & \num{7} & \num{9e+19}$^{\ddagger}$ & \num{9e+19} & \num{0.23} & \num{2.1e+19} & \num{1.4e+20} \\
SPARC & Tokamak & 2027 & Projected & \onlinecite{2022_Wurzel_Hsu} & \num{20} & \num{22} & \num{4e+20}$^{\ddagger}$ & \num{4e+20} & \num{0.77} & \num{3.1e+20} & \num{6.2e+21} \\
ITER & Tokamak & 2039 & Projected & \onlinecite{2022_Wurzel_Hsu} & \num{20} & – & \num{1e+20}$^{\ddagger}$ & \num{1e+20} & \num{3.7} & \num{3.7e+20} & \num{7.4e+21} \\
\noalign{\smallskip}\hline
\end{tabular*}
                              \raggedright
                              \footnotesize{\\$*$ Peak value of density or temperature has been inferred from volume-averaged value as described in Sec.~IV A 4  of the original paper. \cite{2022_Wurzel_Hsu}\\
$\dagger$ Ion temperature has been inferred from electron temperature as described in Sec.~IV A 5 of the original paper. \cite{2022_Wurzel_Hsu}\\
$\ddagger$ Ion density has been inferred from electron density as described in Sec.~IV A 5 of the original paper. \cite{2022_Wurzel_Hsu}\\
$\#$ Energy confinement time $\tau_E^*$ (TFTR/Lawson method) has been inferred from a measurement of the energy confinement time $\tau_E$ (JET/JT-60) method as described in Sec.~IV A 6 of the original paper. \cite{2022_Wurzel_Hsu}}
\end{sidewaystable*}

\begin{sidewaystable*}[p]
\caption{Data for other MCF (i.e., other than tokamaks or spherical tokamaks) and lower-density MIF concepts.}
\label{tab:alternates_mcf_data_table}
\begin{tabular*}{\textwidth}{@{\extracolsep{\fill}}llllllllllll}
\hline\noalign{\smallskip}
Project & Concept & Date & Shot identifier & Reference & \thead{$T_{i0}$ \\ (\si{keV})} & \thead{$T_{e0}$ \\ (\si{keV})} & \thead{$n_{i0}$ \\ (\si{m^{-3}})} & \thead{$n_{e0}$ \\ (\si{m^{-3}})} & \thead{$\tau_{E}^{*}$ \\ (\si{s})} & \thead{$n_{i0} \tau_{E}^{*}$ \\ (\si{m^{-3}~s})} & \thead{$n_{i0} T_{i0} \tau_{E}^{*}$ \\ (\si{keV~m^{-3}~s})} \\
\noalign{\smallskip}\hline\noalign{\smallskip}
ZETA & Pinch & 1957 & 140 kA - 180kA discharges & \onlinecite{2022_Wurzel_Hsu} & \num{0.09} & \num{0.03} & \num{1e+20}$^{\ddagger}$ & \num{1e+20} & \num{0.0001} & \num{1.0e+16} & \num{9.0e+14} \\
Model C & Stellarator & 1969 & ICRH heated & \onlinecite{2022_Wurzel_Hsu} & \num{0.2}$^{\dagger}$ & \num{0.2} & \num{5e+18}$^{\ddagger}$ & \num{5e+18} & \num{0.0001} & \num{5.0e+14} & \num{1.0e+14} \\
ETA-BETA I & RFP & 1977 & Summary & \onlinecite{2022_Wurzel_Hsu} & \num{0.01} & – & \num{1e+21} & – & \num{1e-06} & \num{1.0e+15} & \num{1.0e+13} \\
W7-A & Stellarator & 1980 & Zero current & \onlinecite{2022_Wurzel_Hsu} & \num{0.545} & \num{0.316} & \num{9.6e+19}$^{\ddagger}$ & \num{9.6e+19} & \num{0.0165} & \num{1.6e+18} & \num{8.6e+17} \\
ETA-BETA II & RFP & 1984 & 59611 & \onlinecite{2022_Wurzel_Hsu} & \num{0.09}$^{\dagger}$ & \num{0.09} & \num{3.5e+20}$^{\ddagger}$ & \num{3.5e+20} & \num{0.0001} & \num{3.5e+16} & \num{3.2e+15} \\
TMX-U & Mirror & 1984-02-02 & 2/2/84 S21 & \onlinecite{2022_Wurzel_Hsu} & \num{0.15} & \num{0.045} & \num{2e+18} & \num{2e+18} & \num{0.001} & \num{2.0e+15} & \num{3.0e+14} \\
ZT-40M & RFP & 1987 & 330 kA discharge & \onlinecite{2022_Wurzel_Hsu} & \num{0.33}$^{\dagger}$ & \num{0.33} & \num{9.60e+19}$^{\ddagger *}$ & – & \num{0.0007} & \num{6.7e+16} & \num{2.2e+16} \\
CTX & Spheromak & 1990 & Solid flux conserver & \onlinecite{2022_Wurzel_Hsu} & \num{0.18} & \num{0.18} & \num{4.50e+19}$^{\ddagger *}$ & – & \num{0.0002} & \num{9.0e+15} & \num{1.6e+15} \\
LSX & FRC & 1993 & s~2 & \onlinecite{2022_Wurzel_Hsu} & \num{0.547} & \num{0.253} & \num{1.30e+21}$^*$ & – & \num{0.0001} & \num{1.3e+17} & \num{7.1e+16} \\
MST & RFP & 2001 & 390 kA discharge & \onlinecite{2022_Wurzel_Hsu} & \num{0.396} & \num{0.792} & \num{1.20e+19}$^{\ddagger *}$ & – & \num{0.0064}$^{\#}$ & \num{7.7e+16} & \num{3.0e+16} \\
W7-AS & Stellarator & 2002 & H-NBI mode & \onlinecite{2022_Wurzel_Hsu} & \num{2.28}$^*$ & – & \num{1.10e+20}$^{\ddagger *}$ & – & \num{0.06} & \num{6.6e+18} & \num{1.5e+19} \\
FRX-L & FRC & 2003 & 2027 & \onlinecite{2022_Wurzel_Hsu} & \num{0.09} & \num{0.09} & \num{4e+22} & \num{4e+22} & \num{3.3e-06} & \num{1.3e+17} & \num{1.2e+16} \\
ZaP & Z Pinch & 2003 & Unknown & \onlinecite{2022_Wurzel_Hsu} & \num{0.1} & – & \num{9e+22}$^{\ddagger}$ & \num{9e+22} & \num{3.7e-07} & \num{3.3e+16} & \num{3.3e+15} \\
FRX-L & FRC & 2005 & 3684 & \onlinecite{2022_Wurzel_Hsu} & \num{0.18}$^*$ & – & \num{4.81e+22}$^{\ddagger *}$ & – & \num{3.3e-06} & \num{1.6e+17} & \num{2.9e+16} \\
HSX & Stellarator & 2005 & QHS configuration & \onlinecite{2022_Wurzel_Hsu} & \num{0.45}$^{\dagger}$ & \num{0.45} & \num{2.5e+18}$^{\ddagger}$ & \num{2.5e+18} & \num{0.0006} & \num{1.5e+15} & \num{6.8e+14} \\
TCS & FRC & 2005 & 9018 & \onlinecite{2022_Wurzel_Hsu} & \num{0.025} & \num{0.025} & \num{6.50e+18}$^{\ddagger *}$ & – & \num{4e-05} & \num{2.6e+14} & \num{6.5e+12} \\
SSPX & Spheromak & 2007 & 17524 & \onlinecite{2022_Wurzel_Hsu} & \num{0.5}$^{\dagger}$ & \num{0.5} & \num{2.25e+20}$^{\ddagger *}$ & – & \num{0.001} & \num{2.2e+17} & \num{1.1e+17} \\
GOL-3 & Mirror & 2007 & Unknown & \onlinecite{2022_Wurzel_Hsu} & \num{2} & \num{2} & \num{7e+20} & \num{7e+20} & \num{0.0009} & \num{6.3e+17} & \num{1.3e+18} \\
RFX-mod & RFP & 2008 & 24063 & \onlinecite{2022_Wurzel_Hsu} & \num{1}$^{\dagger}$ & \num{1} & \num{3e+19}$^{\ddagger}$ & \num{3e+19} & \num{0.0025} & \num{7.5e+16} & \num{7.5e+16} \\
TCSU & FRC & 2008 & 21214 & \onlinecite{2022_Wurzel_Hsu} & \num{0.1} & \num{0.1} & \num{1.30e+19}$^{\ddagger *}$ & – & \num{7.5e-05} & \num{9.7e+14} & \num{9.7e+13} \\
LHD & Stellarator & 2008 & High triple product & \onlinecite{2022_Wurzel_Hsu} & \num{0.47} & \num{0.47} & \num{5e+20}$^{\ddagger}$ & \num{5e+20} & \num{0.22} & \num{1.1e+20} & \num{5.2e+19} \\
MST & RFP & 2009 & w/o pellets & \onlinecite{2022_Wurzel_Hsu} & \num{1.3} & \num{1.9} & \num{1.2e+19}$^{\ddagger}$ & \num{1.2e+19} & \num{0.012} & \num{1.4e+17} & \num{1.9e+17} \\
MST & RFP & 2009 & pellets & \onlinecite{2022_Wurzel_Hsu} & \num{0.6} & \num{0.7} & \num{4e+19}$^{\ddagger}$ & \num{4e+19} & \num{0.007} & \num{2.8e+17} & \num{1.7e+17} \\
IPA & FRC & 2010 & Unknown & \onlinecite{2022_Wurzel_Hsu} & \num{0.85}$^*$ & – & \num{5.20e+21}$^{\ddagger *}$ & – & \num{1e-05} & \num{5.2e+16} & \num{4.4e+16} \\
Yingguang-I & FRC & 2015 & 150910-01 & \onlinecite{2022_Wurzel_Hsu} & \num{0.2}$^*$ & – & \num{4.81e+22}$^{\ddagger *}$ & – & \num{1e-06} & \num{4.8e+16} & \num{9.6e+15} \\
C-2U & FRC & 2017 & 46366 & \onlinecite{2022_Wurzel_Hsu} & \num{0.68}$^*$ & – & \num{2.47e+19}$^{\ddagger *}$ & – & \num{0.00024} & \num{5.9e+15} & \num{4.0e+15} \\
W7-X & Stellarator & 2017-12-07 & W7X 20171207.006 & \onlinecite{2022_Wurzel_Hsu} & \num{3.5} & \num{3.5} & \num{8e+19}$^{\ddagger}$ & \num{8e+19} & \num{0.22} & \num{1.8e+19} & \num{6.2e+19} \\
FuZE & Z Pinch & 2018 & Multiple identical shots & \onlinecite{2022_Wurzel_Hsu} & \num{1.8} & – & \num{1.1e+23}$^{\ddagger}$ & \num{1.1e+23} & \num{1.1e-06} & \num{1.2e+17} & \num{2.2e+17} \\
GDT & Mirror & 2018 & Multiple identical shots & \onlinecite{2022_Wurzel_Hsu} & \num{0.45}$^{\dagger}$ & \num{0.45} & \num{1.1e+19}$^{\ddagger}$ & \num{1.1e+19} & \num{0.0006} & \num{6.6e+15} & \num{3.0e+15} \\
PCS & MTF & 2018-06-01 & PCS-16 (post-compression) & \onlinecite{2024_Howard} & \num{0.629} & – & \num{5e+20}$^{\ddagger}$ & \num{5e+20} & \num{0.000122} & \num{6.1e+16} & \num{3.8e+16} \\
PCS & MTF & 2018-06-01 & PCS-16 (pre-compression) & \onlinecite{2024_Howard} & \num{0.600} & \num{0.207} & \num{1.25e+20}$^{\ddagger}$ & \num{1.25e+20} & \num{8.4e-05} & \num{1.0e+16} & \num{6.3e+15} \\
C-2W & FRC & 2019 & 107322 & \onlinecite{2022_Wurzel_Hsu} & \num{0.6}$^*$ & – & \num{2.08e+19}$^{\ddagger *}$ & – & \num{0.0012} & \num{2.5e+16} & \num{1.5e+16} \\
C-2W & FRC & 2019 & 104989 & \onlinecite{2022_Wurzel_Hsu} & \num{1.0}$^*$ & – & \num{1.30e+19}$^{\ddagger *}$ & – & \num{0.003} & \num{3.9e+16} & \num{3.9e+16} \\
C-2W & FRC & 2020 & 114534 & \onlinecite{2022_Wurzel_Hsu} & \num{1.8}$^*$ & – & \num{1.30e+19}$^{\ddagger *}$ & – & \num{0.0015} & \num{2.0e+16} & \num{3.5e+16} \\
C-2W & FRC & 2021 & 118340 & \onlinecite{2022_Wurzel_Hsu} & \num{3.5}$^*$ & – & \num{1.30e+19}$^{\ddagger *}$ & – & \num{0.005} & \num{6.5e+16} & \num{2.3e+17} \\
FuZE & Z Pinch & 2022 & Multiple identical shots at 200psi & \onlinecite{2024_Goyon} & \num{2.4}$^{\dagger}$ & \num{2.4} & \num{4.9e+23}$^{\ddagger}$ & \num{4.9e+23} & \num{1e-06} & \num{4.9e+17} & \num{1.2e+18} \\
C-2W & FRC & 2023 & 122588 & \onlinecite{2024_Gota_Private,2024_Gota} & \num{1}$^{\dagger}$ & \num{1} & \num{1.69e+19}$^{\ddagger *}$ & – & \num{0.002} & \num{3.4e+16} & \num{3.4e+16} \\
\noalign{\smallskip}\hline
\end{tabular*}
                              \raggedright
                              \footnotesize{\\$*$ Peak value of density or temperature has been inferred from volume-averaged value as described in Sec.~IV A 4  of the original paper. \cite{2022_Wurzel_Hsu}\\
$\dagger$ Ion temperature has been inferred from electron temperature as described in Sec.~IV A 5 of the original paper. \cite{2022_Wurzel_Hsu}\\
$\ddagger$ Ion density has been inferred from electron density as described in Sec.~IV A 5 of the original paper. \cite{2022_Wurzel_Hsu}\\
$\#$ Energy confinement time $\tau_E^*$ (TFTR/Lawson method) has been inferred from a measurement of the energy confinement time $\tau_E$ (JET/JT-60) method as described in Sec.~IV A 6 of the original paper. \cite{2022_Wurzel_Hsu}}
\end{sidewaystable*}

\begin{sidewaystable*}[p]
\caption{Data for ICF and higher-density MIF concepts.}
\label{tab:icf_mif_data_table}
\begin{tabular*}{\textwidth}{@{\extracolsep{\fill}}llllllllllllllll}
\hline\noalign{\smallskip}
Project & Concept & Date & Shot identifier & Reference & \thead{$\langle T_i \rangle_{\rm n}$ \\ (\si{keV})} & \thead{$T_e$ \\ (\si{keV})} & \thead{$\rho R_{tot(n)}^{no (\alpha)}$ \\ (\si{g/cm^{-2}})} & YOC & \thead{$p_{stag}$ \\ (\si{Gbar})} & \thead{$\tau_{stag}$ \\ (\si{s})} & \thead{$E_{\rm in}$ \\ (\si{J})} & \thead{$Y$ \\ (\si{J})} & \thead{$P\tau_{\rm stag}$ \\ (\si{atm~s})} & \thead{$n\tau_{\rm stag}$ \\ (\si{m^{-3}~s})} & \thead{$n \langle T \rangle_{\rm n} \tau_{\rm stag}$ \\ (\si{keV~m^{-3}~s})} \\
\noalign{\smallskip}\hline\noalign{\smallskip}
NOVA & Laser Indirect Drive & 1994 & 100 atm fill & \onlinecite{2022_Wurzel_Hsu} & 0.9 & – & – & – & 16 & \num{5e-11} & – & – & \num{0.79} & \num{2.8e+20} & \num{2.5e+20} \\
OMEGA & Laser Direct Drive & 2007-04-17 & 47206 & \onlinecite{2022_Wurzel_Hsu} & 2.0 & – & 0.202 & 0.1 & – & – & – & – & \num{1.23} & \num{1.9e+20} & \num{3.9e+20} \\
OMEGA & Laser Direct Drive & 2007-04-17 & 47210 & \onlinecite{2022_Wurzel_Hsu} & 2.0 & – & 0.182 & 0.1 & – & – & – & – & \num{1.13} & \num{1.8e+20} & \num{3.6e+20} \\
OMEGA & Laser Direct Drive & 2009 & Unknown & \onlinecite{2022_Wurzel_Hsu} & 1.8 & – & 0.240 & 0.1 & – & – & – & – & \num{1.29} & \num{2.3e+20} & \num{4.1e+20} \\
OMEGA & Laser Direct Drive & 2009-09-09 & 55468 & \onlinecite{2022_Wurzel_Hsu} & 1.8 & – & 0.300 & 0.1 & – & – & – & – & \num{1.55} & \num{2.7e+20} & \num{4.9e+20} \\
OMEGA & Laser Direct Drive & 2013-04-03 & 69236 & \onlinecite{2022_Wurzel_Hsu} & 2.8 & – & – & – & 18 & \num{1.15e-10} & – & – & \num{2.04} & \num{2.3e+20} & \num{6.5e+20} \\
NIF & Laser Indirect Drive & 2014-03-04 & N140304 & \onlinecite{2022_Wurzel_Hsu} & 5.5 & – & – & – & 222 & \num{1.63e-10} & – & – & \num{35.71} & \num{2.1e+21} & \num{1.1e+22} \\
Z Facility & MagLIF & 2014-03-05 & z2613 & \onlinecite{2022_Wurzel_Hsu} & 2.0 & – & – & – & 0.56 & \num{1.38e-09} & – & – & \num{0.76} & \num{1.2e+20} & \num{2.4e+20} \\
OMEGA & Laser Direct Drive & 2015-04-28 & 77068 & \onlinecite{2022_Wurzel_Hsu} & 3.6 & – & – & – & 56 & \num{6.6e-11} & – & – & \num{3.65} & \num{3.2e+20} & \num{1.2e+21} \\
Z Facility & MagLIF & 2015-09-15 & z2850 & \onlinecite{2022_Wurzel_Hsu} & 2.8 & – & – & – & 0.6 & \num{1.62e-09} & – & – & \num{0.96} & \num{1.1e+20} & \num{3.0e+20} \\
NIF & Laser Indirect Drive & 2017-06-01 & N170601 & \onlinecite{2022_Wurzel_Hsu} & 4.5 & – & – & – & 320 & \num{1.6e-10} & \num{1.5e+06} & \num{4.8e+04} & \num{50.53} & \num{3.6e+21} & \num{1.6e+22} \\
NIF & Laser Indirect Drive & 2017-08-27 & N170827 & \onlinecite{2022_Wurzel_Hsu} & 4.5 & – & – & – & 360 & \num{1.54e-10} & \num{1.7e+06} & \num{5.3e+04} & \num{54.72} & \num{3.8e+21} & \num{1.7e+22} \\
Z Facility & MagLIF & 2017-12-04 & z3179 & \onlinecite{2022_Knapp} & 3.3 & – & – & – & 1.2 & \num{1.8e-09} & – & – & \num{2.13} & \num{2.0e+20} & \num{6.7e+20} \\
Z Facility & MagLIF & 2018-04-02 & z3236 & \onlinecite{2022_Knapp} & 2.9 & – & – & – & 1.3 & \num{2.1e-09} & – & – & \num{2.69} & \num{2.9e+20} & \num{8.5e+20} \\
Z Facility & MagLIF & 2018-08-15 & z3289 & \onlinecite{2022_Knapp} & 2.6 & – & – & – & 1.8 & \num{2e-09} & – & – & \num{3.55} & \num{4.3e+20} & \num{1.1e+21} \\
FIREX & Laser Direct Drive & 2019 & 40558 & \onlinecite{2022_Wurzel_Hsu} & – & 2.1 & – & – & 2 & \num{4e-10} & – & – & \num{0.79} & \num{1.2e+20} & \num{2.5e+20} \\
NIF & Laser Indirect Drive & 2019-09-18 & N190918 & \onlinecite{Zylstra_2021} & 4.43 & – & – & – & 140 & \num{1.54e-10} & \num{1.9e+06} & \num{2.1e+04} & \num{21.28} & \num{1.5e+21} & \num{6.7e+21} \\
NIF & Laser Indirect Drive & 2019-10-07 & N191007 & \onlinecite{2022_Wurzel_Hsu} & 4.52 & – & – & – & 206 & \num{1.51e-10} & \num{1.9e+06} & \num{5.3e+04} & \num{30.70} & \num{2.1e+21} & \num{9.7e+21} \\
NIF & Laser Indirect Drive & 2019-11-10 & N191110 & \onlinecite{Zylstra_2021} & 4.54 & – & – & – & 169 & \num{1.66e-10} & \num{1.9e+06} & \num{5.6e+04} & \num{27.69} & \num{1.9e+21} & \num{8.8e+21} \\
NIF & Laser Indirect Drive & 2020-11-01 & N201101 & \onlinecite{2022_Wurzel_Hsu} & 4.61 & – & – & – & 319 & \num{1.18e-10} & – & – & \num{37.15} & \num{2.5e+21} & \num{1.2e+22} \\
NIF & Laser Indirect Drive & 2020-11-22 & N201122 & \onlinecite{2022_Wurzel_Hsu} & 4.65 & – & – & – & 297 & \num{1.37e-10} & – & – & \num{40.16} & \num{2.7e+21} & \num{1.3e+22} \\
NIF & Laser Indirect Drive & 2021-02-07 & N210207 & \onlinecite{2024_Abu-Shawareb} & 5.66 & – & – & – & 304 & \num{1.37e-10} & \num{1.9e+06} & \num{1.7e+05} & \num{41.10} & \num{2.3e+21} & \num{1.3e+22} \\
NIF & Laser Indirect Drive & 2021-02-20 & N210220 & \onlinecite{2022_Wurzel_Hsu} & 5.13 & – & – & – & 371 & \num{1.35e-10} & – & – & \num{49.43} & \num{3.0e+21} & \num{1.6e+22} \\
NIF & Laser Indirect Drive & 2021-03-07 & N210307 & \onlinecite{2024_Abu-Shawareb} & 5.55 & – & – & – & 323 & \num{1.38e-10} & \num{1.9e+06} & \num{1.4e+05} & \num{43.99} & \num{2.5e+21} & \num{1.4e+22} \\
NIF & Laser Indirect Drive & 2021-08-08 & N210808 & \onlinecite{2024_Abu-Shawareb} & 10.9 & – & – & – & 561 & \num{8.9e-11} & \num{1.9e+06} & \num{1.3e+06} & \num{49.28} & \num{1.4e+21} & \num{1.6e+22} \\
OMEGA & Laser Direct Drive & 2021-10-14 & 102154 & \onlinecite{2024_Williams} & 4.6 & – & – & – & 76.3 & \num{6e-11} & \num{3.0e+04} & \num{6.3e+02} & \num{4.52} & \num{3.1e+20} & \num{1.4e+21} \\
OMEGA & Laser Direct Drive & 2021-11-02 & 102360 & \onlinecite{2024_Williams} & 6.0 & – & – & – & 34.9 & \num{6e-11} & \num{3.0e+04} & \num{7.5e+02} & \num{2.07} & \num{1.1e+20} & \num{6.5e+20} \\
OMEGA & Laser Direct Drive & 2022-04-14 & 103952 & \onlinecite{2024_Williams} & 5.8 & – & – & – & 46.2 & \num{6e-11} & \num{3.0e+04} & \num{8.7e+02} & \num{2.74} & \num{1.5e+20} & \num{8.7e+20} \\
OMEGA & Laser Direct Drive & 2022-07-14 & 104949 & \onlinecite{2024_Gopalaswamy} & 4.6 & 3.8 & 0.160 & – & 78 & \num{7e-11} & \num{3.0e+04} & \num{5.9e+02} & \num{5.39} & \num{3.7e+20} & \num{1.7e+21} \\
\noalign{\smallskip}\hline
\end{tabular*}
\end{sidewaystable*}

\begin{sidewaystable*}[p]
\caption{Data for experiments that produced sufficient fusion energy to achieve appreciable values of scientific energy gain $Q_{\mathrm{sci}}$.}
\label{tab:q_sci_data_table}
\begin{tabular*}{\textwidth}{@{\extracolsep{\fill}}llllllllll}
\hline\noalign{\smallskip}
Project & Concept & Date & Shot identifier & Reference & \thead{$E_{\rm in}$ \\ (\si{J})} & \thead{$Y$ \\ (\si{J})} & \thead{$P_{\rm in}$ \\ (\si{W})} & \thead{$P_{\rm F}$ \\ (\si{W})} & \thead{$Q_{\rm sci}$ \\ } \\
\noalign{\smallskip}\hline\noalign{\smallskip}
JET & Tokamak & 1991-11-09 & 26148 & \onlinecite{2022_Wurzel_Hsu} & – & – & \num{1.4e+07} & \num{1.7e+06} & 0.12 \\
TFTR & Tokamak & 1992-10-29 & 68522 & \onlinecite{2022_Wurzel_Hsu} & – & – & \num{3.1e+07} & \num{6.5e+04} & 0.00 \\
TFTR & Tokamak & 1994-05-27 & 76778 & \onlinecite{2022_Wurzel_Hsu} & – & – & \num{3.4e+07} & \num{9.3e+06} & 0.28 \\
TFTR & Tokamak & 1994-11-02 & 80539 & \onlinecite{2022_Wurzel_Hsu} & – & – & \num{4.0e+07} & \num{1.1e+07} & 0.27 \\
TFTR & Tokamak & 1995-02-17 & 83546 & \onlinecite{2022_Wurzel_Hsu} & – & – & \num{1.7e+07} & \num{2.8e+06} & 0.16 \\
JET & Tokamak & 1997-10-31 & 42974 & \onlinecite{Keilhacker_1999} & – & – & \num{2.6e+07} & \num{1.6e+07} & 0.62 \\
JET & Tokamak & 1997-10-31 & 42976 & \onlinecite{2022_Wurzel_Hsu} & – & \num{1.4e+07} & \num{2.6e+07} & \num{1.6e+07} & 0.63 \\
NIF & Laser Indirect Drive & 2013-09-27 & N130927 & \onlinecite{2014_Hurricane} & \num{1.8e+06} & \num{1.4e+04} & – & – & 0.01 \\
NIF & Laser Indirect Drive & 2013-11-19 & N131119 & \onlinecite{2014_Hurricane} & \num{1.9e+06} & \num{1.7e+04} & – & – & 0.01 \\
NIF & Laser Indirect Drive & 2017-06-01 & N170601 & \onlinecite{2022_Wurzel_Hsu} & \num{1.5e+06} & \num{4.8e+04} & – & – & 0.03 \\
NIF & Laser Indirect Drive & 2017-08-27 & N170827 & \onlinecite{2022_Wurzel_Hsu} & \num{1.7e+06} & \num{5.3e+04} & – & – & 0.03 \\
NIF & Laser Indirect Drive & 2019-09-18 & N190918 & \onlinecite{Zylstra_2021} & \num{1.9e+06} & \num{2.1e+04} & – & – & 0.01 \\
NIF & Laser Indirect Drive & 2019-10-07 & N191007 & \onlinecite{2022_Wurzel_Hsu} & \num{1.9e+06} & \num{5.3e+04} & – & – & 0.03 \\
NIF & Laser Indirect Drive & 2019-11-10 & N191110 & \onlinecite{Zylstra_2021} & \num{1.9e+06} & \num{5.6e+04} & – & – & 0.03 \\
NIF & Laser Indirect Drive & 2021-02-07 & N210207 & \onlinecite{2024_Abu-Shawareb} & \num{1.9e+06} & \num{1.7e+05} & – & – & 0.09 \\
NIF & Laser Indirect Drive & 2021-03-07 & N210307 & \onlinecite{2024_Abu-Shawareb} & \num{1.9e+06} & \num{1.4e+05} & – & – & 0.07 \\
NIF & Laser Indirect Drive & 2021-08-08 & N210808 & \onlinecite{2024_Abu-Shawareb} & \num{1.9e+06} & \num{1.3e+06} & – & – & 0.69 \\
OMEGA & Laser Direct Drive & 2021-10-14 & 102154 & \onlinecite{2024_Williams} & \num{3.0e+04} & \num{6.3e+02} & – & – & 0.02 \\
OMEGA & Laser Direct Drive & 2021-11-02 & 102360 & \onlinecite{2024_Williams} & \num{3.0e+04} & \num{7.5e+02} & – & – & 0.02 \\
JET & Tokamak & 2021-12-21 & 99972 & \onlinecite{2023_Maslov} & – & \num{5.6e+07} & \num{3.3e+07} & \num{1.2e+07} & 0.38 \\
JET & Tokamak & 2021-12-21 & 99971 & \onlinecite{2023_Maslov} & – & \num{5.9e+07} & \num{3.1e+07} & \num{1.0e+07} & 0.33 \\
OMEGA & Laser Direct Drive & 2022-04-14 & 103952 & \onlinecite{2024_Williams} & \num{3.0e+04} & \num{8.7e+02} & – & – & 0.03 \\
OMEGA & Laser Direct Drive & 2022-07-14 & 104949 & \onlinecite{2024_Gopalaswamy} & \num{3.0e+04} & \num{5.9e+02} & – & – & 0.02 \\
NIF & Laser Indirect Drive & 2022-09-19 & N220919 & \onlinecite{2024_Abu-Shawareb} & \num{2.0e+06} & \num{1.2e+06} & – & – & 0.59 \\
NIF & Laser Indirect Drive & 2022-12-05 & N221204 & \onlinecite{2024_Abu-Shawareb} & \num{2.0e+06} & \num{3.1e+06} & – & – & 1.51 \\
NIF & Laser Indirect Drive & 2023-07-30 & N230729 & \onlinecite{2025_LLNL} & \num{2.0e+06} & \num{3.9e+06} & – & – & 1.89 \\
JET & Tokamak & 2023-10-03 & 104522 & \onlinecite{2025_Kappatou} & – & \num{6.9e+07} & \num{3.5e+07} & \num{1.3e+07} & 0.37 \\
NIF & Laser Indirect Drive & 2023-10-08 & N231007 & \onlinecite{2025_LLNL} & \num{1.9e+06} & \num{2.4e+06} & – & – & 1.26 \\
JET & Tokamak & 2023-10-09 & 104600 & \onlinecite{2025_Kappatou,2024_Carvalho} & – & \num{2.7e+07} & \num{3.5e+07} & \num{4.0e+06} & 0.11 \\
NIF & Laser Indirect Drive & 2023-10-30 & N231029 & \onlinecite{2025_LLNL} & \num{2.2e+06} & \num{3.4e+06} & – & – & 1.55 \\
NIF & Laser Indirect Drive & 2024-02-12 & N240210 & \onlinecite{2025_LLNL} & \num{2.2e+06} & \num{5.2e+06} & – & – & 2.36 \\
NIF & Laser Indirect Drive & 2024-11-18 & N241117 & \onlinecite{2025_LLNL} & \num{2.2e+06} & \num{4.1e+06} & – & – & 1.86 \\
NIF & Laser Indirect Drive & 2025-02-23 & N250222 & \onlinecite{2025_LLNL} & \num{2.0e+06} & \num{5.0e+06} & – & – & 2.44 \\
NIF & Laser Indirect Drive & 2025-04-07 & N250406 & \onlinecite{2025_LLNL} & \num{2.1e+06} & \num{8.6e+06} & – & – & 4.13 \\
NIF & Laser Indirect Drive & 2025-06-22 & N250622 & \onlinecite{2025_LLNL} & \num{2.1e+06} & \num{2.4e+06} & – & – & 1.17 \\
\noalign{\smallskip}\hline
\end{tabular*}
\end{sidewaystable*}

\begin{acknowledgments}
We acknowledge Julia Haack and Luigi Di Pace for alerting us to the ambiguity in Eq.~(23) of the original paper. Thanks to Mikhail Maslov for providing the dates of specific JET shots, Rich Hawryluk for providing the dates of TFTR shots and feedback on nomenclature, Myles Hildebrand for providing further details on PCS and PI3, Hiroshi Gota for providing further details on recent C-2W shots, Varchas Gopalaswamy for providing the dates of specific OMEGA shots, Matthew Gomez for providing the dates of specific MagLIF shots, both
Leland Ellison and Alison Christopherson for discussions about the adjustment of the confinement time and the requirements for ignition in ICF experiments, and Tammy Ma for providing further details on recent NIF shots.
\end{acknowledgments}

\section*{Data Availability Statement}

Data sharing is not applicable to this paper as no new data were created or analyzed in this study. The extracted data from other publications and codes used to make the plots and tables in this paper are available for download.\cite{Wurzel_2022}

\section*{Author Declarations}
The authors have financial interests in some companies whose experiments appear in the figures and tables of this paper.


\bibliographystyle{aipnum4-1}
\bibliography{lawson-paper-update} 
\end{document}